\documentclass[3p,times]{elsarticle}

\usepackage{ecrc}
\usepackage{url}

\volume{00}

\firstpage{1}

\journalname{Nuclear Physics A}

\runauth{}

\jid{nupha}

\usepackage{amssymb}

\usepackage[figuresright]{rotating}

\newcommand{\pT}{\mbox{$p_\mathrm{T}$}}

\newcommand{\ET}{\mbox{$E_{\mathrm{T}}$}}

\newcommand{\meanTAA}{\mbox{$\langle T_{\mathrm{AA}} \rangle$}}

\newcommand{\Lint}{\mbox{$L_{\mathrm{int}}$}}
\newcommand{\Nevt}{N_{\mathrm{evt}}}

\newcommand{\Nsig}{N^{\mathrm{sig}}}

\newcommand{\Riso}{\mbox{$R_{\mathrm{iso}}$}}
\newcommand{\efftot}{\mbox{$\epsilon_{\mathrm{tot}}$}}

\begin{document}

\begin{frontmatter}

\dochead{}

\title{Measurement of high $\pT$ isolated prompt photons in lead-lead collisions at $\sqrt{s_{\mathrm{NN}} }=2.76$~TeV with the ATLAS detector at the LHC }

\author{}

\address{}

\begin{abstract}
Prompt photons are a powerful tool to study heavy ion collisions.  Their production rates provide access to the initial state parton distribution functions 
and also provide a means to calibrate the expected energy of jets that are produced in the medium.
The ATLAS detector measures photons with its hermetic, longitudinally segmented calorimeter, which gives excellent spatial and energy resolutions, and detailed information about the shower shape of each measured photon.  This provides significant rejection against the expected background from 
the decays of neutral pions in jets.  Rejection against jet fragmentation products is further enhanced by requiring candidate photons to be isolated.  First results on the spectra of isolated prompt photons from a dataset with an 
integrated luminosity of approximately 0.13 nb$^{-1}$ of lead-lead collisions
at $\sqrt{s_{\mathrm{NN}}}=2.76$ TeV are shown as a function of transverse momentum and centrality.  The measured spectra are compared to expectations from perturbative QCD calculations.
\end{abstract}


\end{frontmatter}


\section{Introduction}

While di-jet asymmetry distributions at the Large Hadron Collider 
(LHC)~\cite{Aad:2010bu} are consistent with 
energy loss in the hot, dense medium, the detailed
physical mechanism of ``jet quenching'' 
is still not understood.  One of the limiting factors
in understanding jet quenching is having a proper calibration of the initial jet energy.
Replacing one of the jets with a penetrating
probe, such as a photon or electroweak boson ($W$ or $Z$), offers the possibility of
calibrating the energy of the initial jet.  This was first proposed by Wang and collaborators in Ref.~\cite{Wang:1996yh}.  

Prompt photons are themselves expected to arise via direct emission and fragmentation from jets.
In order to remove background events from di-jet processes, as well as the fragmentation photons,
an ``isolation'' criterion is typically applied 
within a cone of a well-defined radius relative to the photon direction.
In lower-energy heavy ion collisions, 
the PHENIX experiment performed  measurements of direct photon rates in gold-gold collisions at 
$\sqrt{s_{\mathrm{NN}}} = 200$ GeV~\cite{Adler:2005ig}.
The CMS experiment 
performed the first measurement of isolated prompt photon rates in both proton-proton as well as lead-lead collisions,
at the nucleon-nucleon center-of-mass energy of 
$\sqrt{s_{\mathrm{NN}}}=2.76$ TeV~\cite{Chatrchyan:2012vq}.
They found rates consistent with a scaling with the number of binary 
collisions.

This work presents the measurement of the 
yield of prompt isolated photons in the transverse momentum range 
$\pT=45-200$ GeV as a function of collision centrality from the
ATLAS experiment.
The results are described in more detail in Ref.~\cite{HIPhoton}.

\section{Experimental setup and photon reconstruction}

The ATLAS detector is described in detail in Ref.~\cite{Aad:2008zzm}.  
The ATLAS inner detector is comprised of 
three major subsystems: the pixel detector, the semiconductor detector (SCT)
and the transition radiation tracker (TRT), which cover full azimuth 
and pseudorapidity out to $|\eta|=2.5$.
The ATLAS calorimeter is a large-acceptance, longitudinally-segmented sampling calorimeter covering 
$|\eta|<4.9$ with electromagnetic and hadronic sections.  
In test beam environments and in typical proton-proton collisions, 
the calorimeter is found to have a sampling term for electromagnetic showers
of 10-17\%$/\sqrt{E[\mathrm{GeV}]}$.
The total material in front of the electromagnetic 
calorimeter ranges from 2.5 to 6 radiation lengths as a function of pseudorapidity,
except the transition region between the barrel and endcap regions, in which the material is up to 11.5 radiation
lengths.  
The hadronic calorimeter section is located radially just after the electromagnetic calorimeter.  
Within $|\eta|<1.7$, it is a sampling calorimeter of steel and scintillator tiles, with a depth of 
7.4 hadronic interaction lengths.  In the endcap region it is copper and liquid argon with a depth of
9 interaction lengths.

To reconstruct photons in the context of a heavy ion collision, the large background from the underlying event (UE) is subtracted from each event.
This is performed during the heavy ion jet reconstruction, 
which precedes the photon reconstruction,
as explained in detail in Ref.\cite{:2012is}.
This procedure provides a new set of ``subtracted'' cells, from which the mean underlying event, as well as the large-scale modulation from elliptic flow, 
has been removed. 
Acting upon the subtracted cells,
the ATLAS photon reconstruction~\cite{Aad:2009wy} is seeded by clusters of at least 2.5 GeV found using a sliding window algorithm applied 
to the second sampling layer of the electromagnetic calorimeter, which
typically absorbs over 50\% of the deposited photon energy.
The energy measurement is made using all three layers of the electromagnetic calorimeter and the presampler, 
with a size in the barrel region of 
$\Delta\eta\times\Delta\phi = 0.075\times0.0125$.
In the dense environment of the heavy ion collision, the photon 
conversion recovery procedure is not performed, due to the overwhelming number of combinatoric pairs in more central collisions.

To reject clusters arising from hadronic fragments of jets, particularly neutral mesons, the calorimeter is also used to measure
an isolation energy for each photon candidate,
$\ET(\Riso)$.
The isolation energy is the sum of transverse energies in calorimeter cells 
(including hadronic and electromagnetic sections) in a 
cone defined by $R=\sqrt{\Delta \eta^2 + \Delta \phi^2} < \Riso$ around the photon axis, excluding
central core of cells in a region with a size 
corresponding to 5$\times$7 second layer cells.

\section{Data and simulation samples}

The data sample analyzed here is from 
the 2011 LHC heavy ion run, colliding lead nuclei at $\sqrt{s_{\mathrm{NN}}}=2.76$ TeV.
After a trigger requiring a
16 GeV energy deposition in the electromagnetic calorimeter, 
which is sensitive in particular to
both electrons and photons, events were selected which contained
a reconstructed photon or electron candidate 
with a cluster transverse energy of at least 40 GeV.
Events were then further analyzed if they satisfied a set of quality cuts:
the event had to be taken during a period when the detector was found to be working properly,
leaving an integrated luminoisty of approximately $\Lint = 0.13$ nb$^{-1}$ for this analysis, 
after the exclusion of several runs. 
Two sets of Minimum Bias Trigger Scintillators covering $2.09<|\eta|<3.84$ 
are required to have a well reconstructed time signal,
and a relative time between the two counters of less than 5 ns.
Finally, a good vertex is required to be reconstructed in the ATLAS detector, to reject background events 
from e.g. cosmic rays.
In this measurement, yields are presented per minimum bias collision, which are estimated using the total integrated luminosity,
measured using the ATLAS luminosity detectors~\cite{ATLASlumi}.

The centrality of each heavy ion collision is determined using the sum of the 
transverse energy in all cells in the forward calorimeter ($3.1<|\eta|<4.9$), 
at the electromagnetic scale. 
The minimum bias 
trigger and event selection criteria were studied in detail in the 2010 data 
sample~\cite{ATLAS:2011ah} and it was found 
that $98\pm2\%$ of the total inelastic cross section is sampled by the
trigger and event selection requirements.
For this analysis, the data have been divided into four centrality intervals, covering the 0-10\%, 10-20\%, 20-40\% and 40-80\% most central events.
In this convention, the 0-10\% interval has the highest multiplicities, and the 40-80\% the lowest.

For the extraction of photon performance parameters (efficiencies, photon energy scale, isolation properties),
a set of 450,000 photon+jet events generated using the ATLAS MC11 tune of PYTHIA 6.4~\cite{Sjostrand:2006za} at $\sqrt{s}=2.76$ TeV, 
is overlaid on minimum-bias HIJING~\cite{Wang:1991hta} events,
which are referred-to as the ``PYTHIA+HIJING'' sample.
The generated events are fully simulated using GEANT4~\cite{Agostinelli:2002hh} and digitized to produce simulated raw data files, that are reconstructed
and analyzed exactly as is done for experimental data.

\section{Photon selection and reconstruction performance}

\label{section:photoncuts}

The fine-grained, longitudinally segmented calorimeter 
utilized in ATLAS allows detailed characterization of the shape of each photon shower, providing 
tools to reject jets and hadrons, while maintaining high efficiency for the photons themselves.
In this analysis, nine shower-shape variables are used, 
all of which have been used extensively in previous ATLAS measurements, 
particularly the measurement of prompt photon spectra as a function of pseudorapidity~\cite{Aad:2010sp,Aad:2011tw}.
The cuts used in this analysis are ``HI tight'' cuts, defined in Ref.~\cite{HIPhoton} as a minimal set of changes
to the standard set of tight cuts used for unconverted photons in the proton-proton analysis.
Photons are also restricted to be in the pseudorapidity interval $|\eta|<1.3$.

Relative to the energy calibration determined in $pp$ collisions~\cite{Aad:2009wy}, 
the residual correction needed to compensate for the mixture of converted and unconverted photons 
is found to be generally less than a percent for much of the measured 
$\pT$ range.
It was found in the PYTHIA+HIJING sample, 
that there are no significant differences
in the residual corrections between peripheral and central events.

\begin{figure}[t]
\begin{center}
\includegraphics[width=0.4\textwidth]{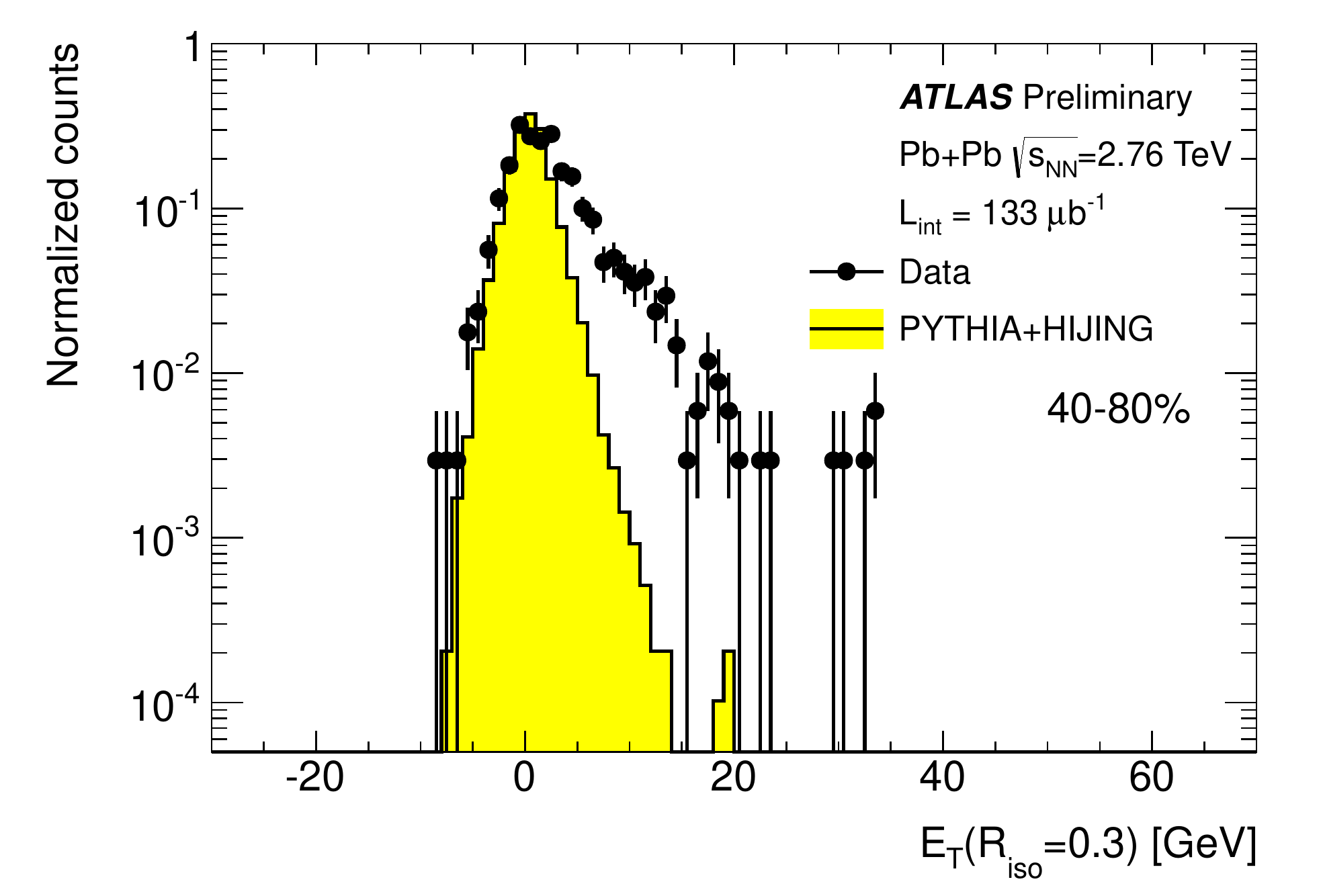}
\includegraphics[width=0.4\textwidth]{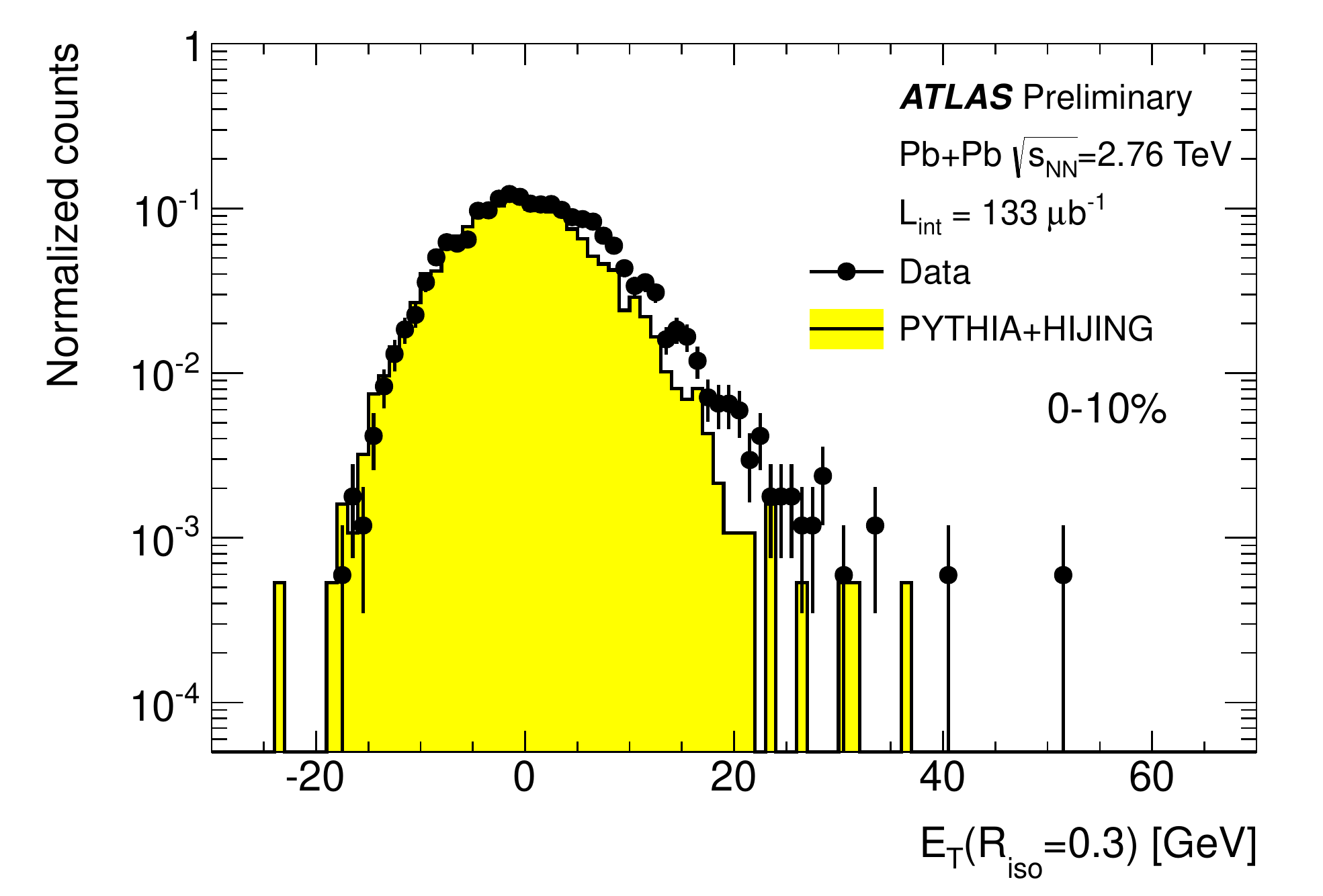}

\caption{Distributions of photon isolation energy in a $\Riso = 0.3$ cone for the 
most peripheral and central centrality
bins in data (black points) and for MC (yellow histogram), normalized for negative
$\ET(\Riso = 0.3)$ values~\cite{HIPhoton}.  
The differences at large values of $\ET(\Riso = 0.3)$ can be attributed to the presence of
jet contamination in the data, which is not present in the MC sample.
}
\label{figure:comp_etcone}
\end{center}
\end{figure}

Figure~\ref{figure:comp_etcone} shows the distributions of the 
isolation energy $\ET(\Riso = 0.3)$ for two centrality bins,
compared with simulated distributions, normalized in the 
region $\ET(\Riso = 0.3)<0$.
Both data and MC distributions grow noticeably wider with increasing centrality, and agree well in the 
region of negative isolation energy.
The observed enhancement on the $\ET(\Riso = 0.3)>0$ side of these distributions is expected from di-jet background events. 

\section{Measurement of photon yields}

\begin{figure}[t]
\begin{center}
\includegraphics[width=0.4\textwidth]{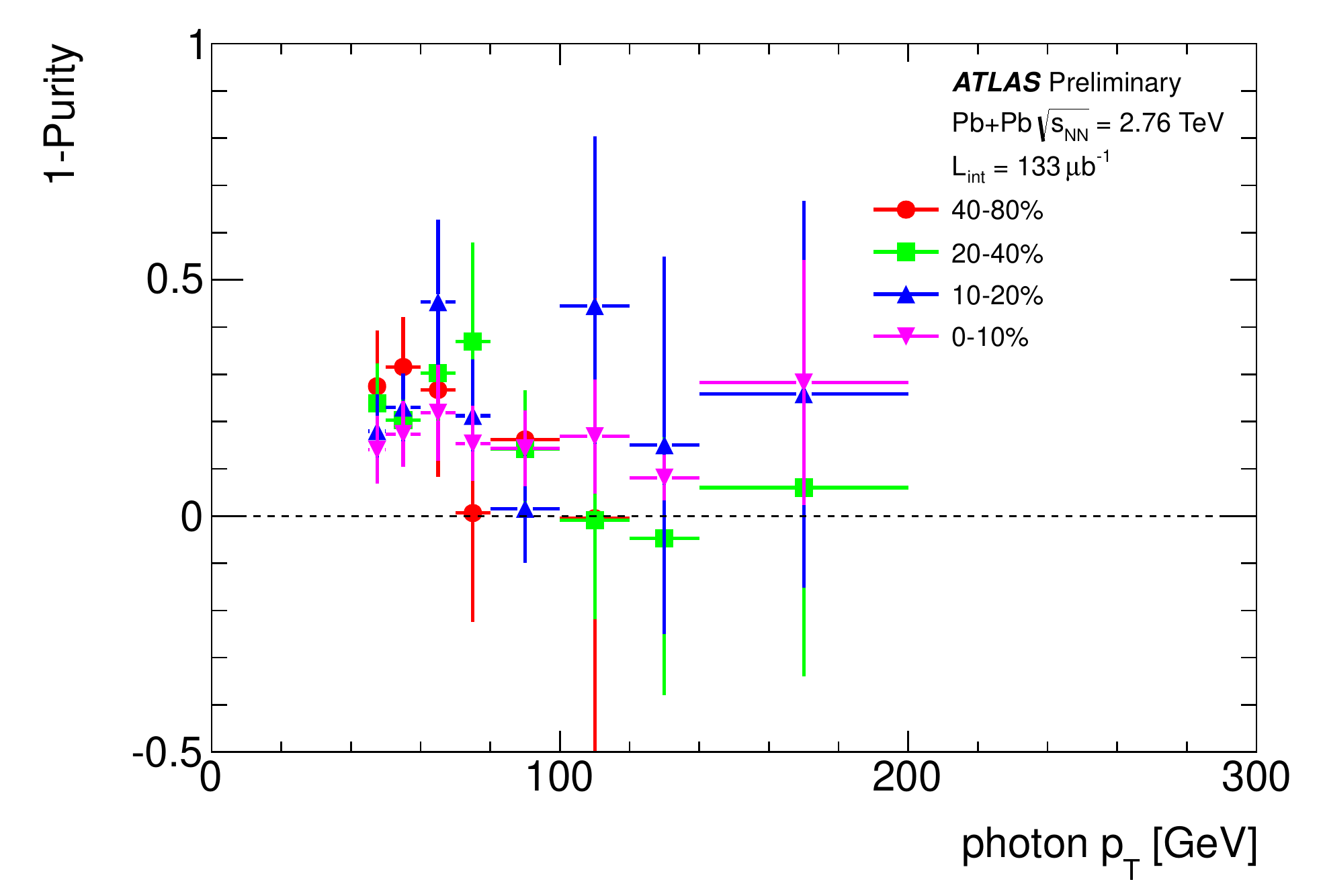}
\includegraphics[width=0.4\textwidth]{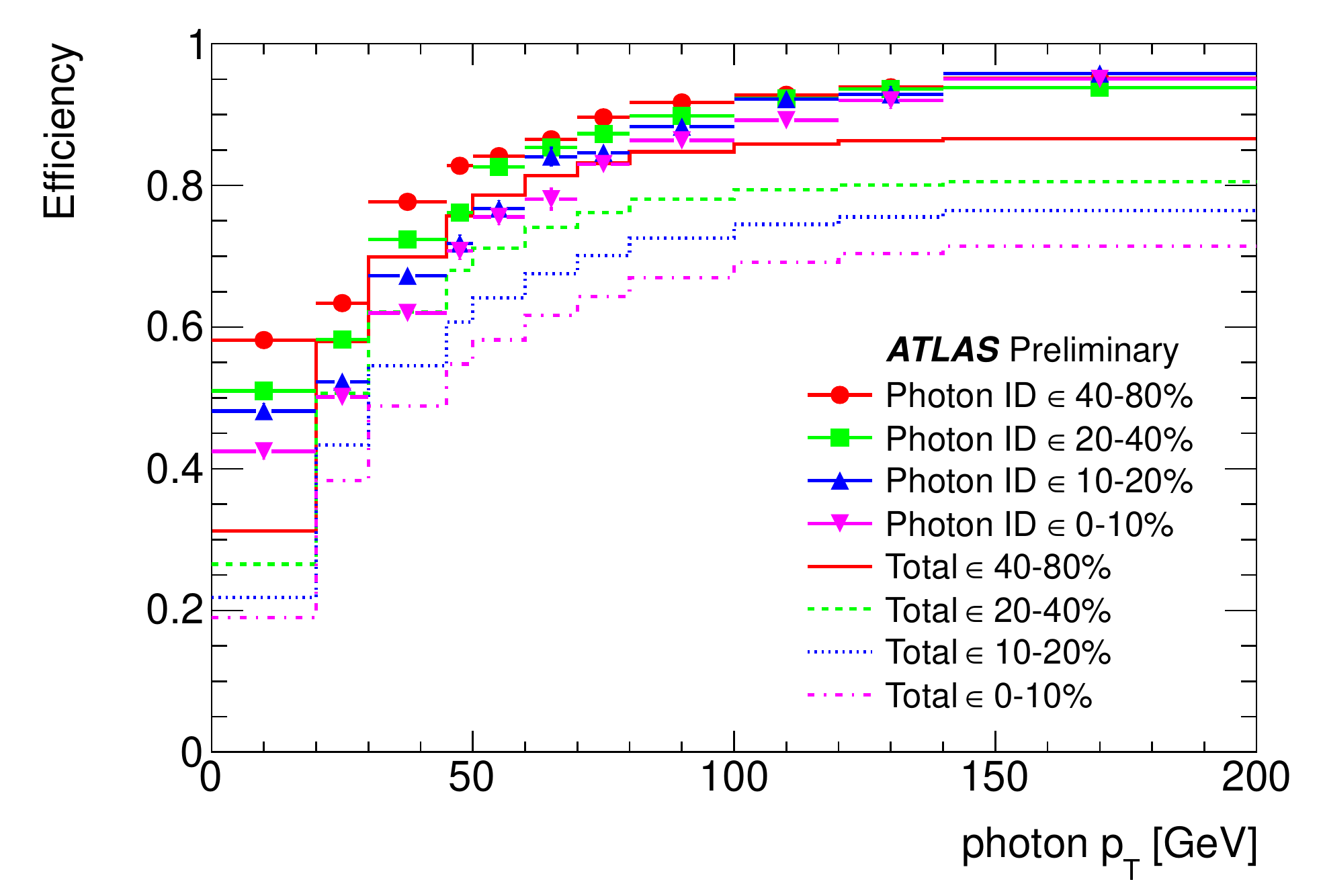}
\caption{
\label{figure:efficiency}
(left) The factor $1-P$ (where $P$ is the purity) extracted from data in each $\pT$ interval for four centrality bins, using the double sideband method~\cite{HIPhoton}.
(right) Photon reconstruction efficiency as a function of photon $\pT$ and event centrality 
averaged over $|\eta|<1.3$, based on MC calculations performed in four subintervals in $\eta$.
The points indicate the identification efficiency, 
without the contribution from the isolation requirement.  The solid lines indicate the smoothed
total efficiency used to correct the measured photon yield~\cite{HIPhoton}.
}
\end{center}
\end{figure}

Di-jet backgrounds are subtracted using the so-called ``double sideband method''~\cite{Aad:2010sp,Aad:2011tw})
where photons are binned along two axes:
the isolation energy within a chosen isolation cone size,
and the photon selection criterion (``tight'' or ``non tight'').
Non-tight photons fail at least one of the more stringent cuts, and have an enhanced di-jet contribution.
The double sideband approach uses the ratio of counts for non-tight photons which satisfy or fail the isolation requirement
to extrapolate the measured number of tight non-isolated photons into the signal region.
The purity extracted using this technique is shown in the left panel of
Figure~\ref{figure:efficiency} (as $1-$Purity) for four centrality intervals.

The final conversion of the measured yield into a yield per event
requires 
two more factors:
the total number of events in each centrality sample, and a reconstruction efficiency including all known effects.
The efficiencies are defined for ``HI tight'', isolated photons relative to all PYTHIA photons with an isolation energy in a cone
of $\Riso = 0.3$ around the photon direction of less than 6 GeV.
The needed efficiency corrections are categorized into three broad classes:
reconstruction efficiency, 
identification efficiency, and 
isolation efficiency. 
These are defined in such a way that the ``total efficiency'' $\efftot$ 
is simply the product of these three factors.
The right panel of 
Figure~\ref{figure:efficiency} shows the product of the reconstruction and 
identification efficiency (points) and the total efficiency (solid lines).

Several sources of systematic uncertainty have been estimated by varying the assumptions applied to the analysis:
the definition of tight and non-tight cuts, the event counting procedure, the photon energy scale, and the photon 
energy resolution.  
The uncertainties are dominated by the variations in the choice of 
the photon identification cuts and the
precise choice of isolation cone properties, and a $\pT$- and centrality-independent uncertainty of 31\%
is applied.

\begin{figure}[t]
\begin{center}
\includegraphics[width=0.4\textwidth]{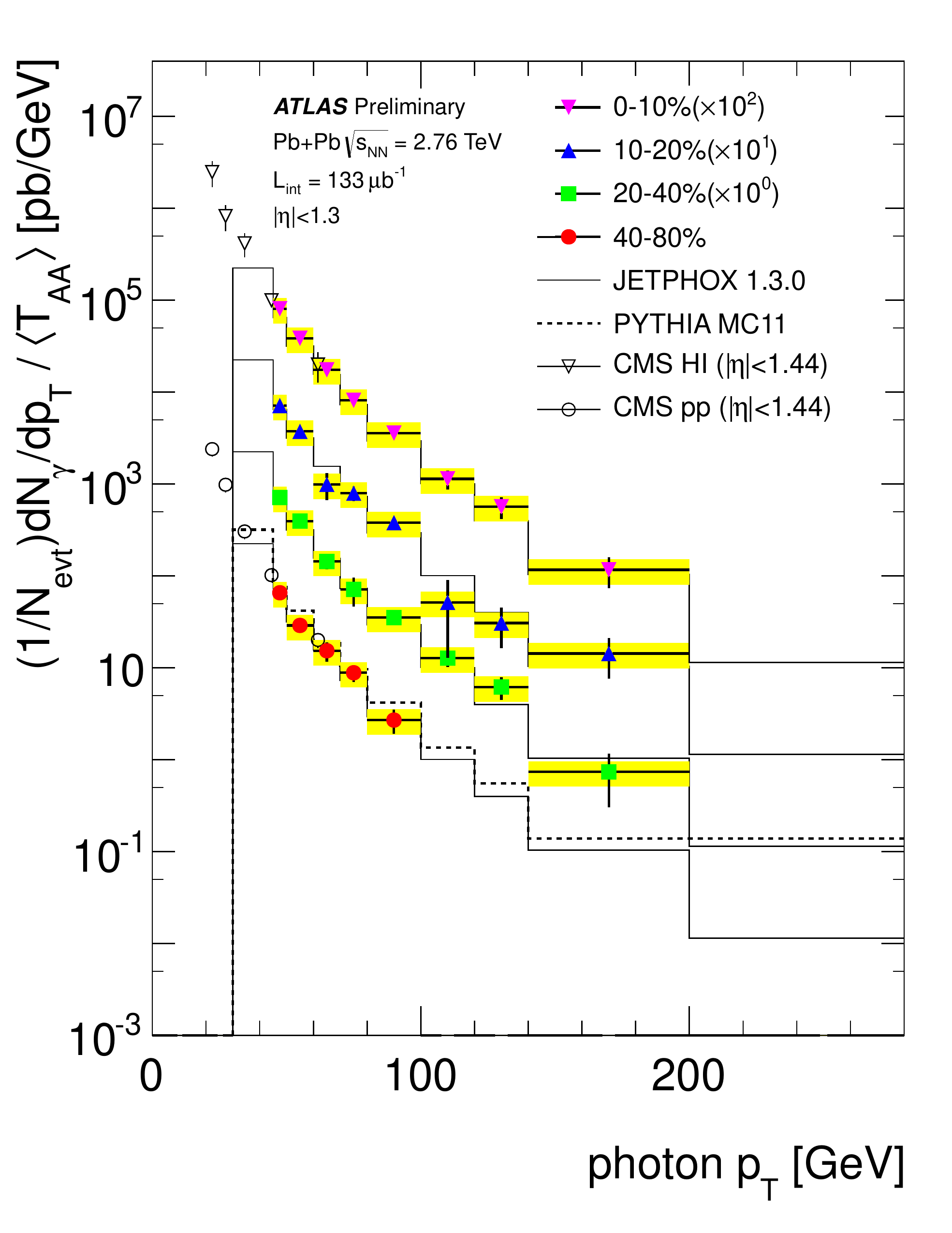}
\includegraphics[width=0.4\textwidth]{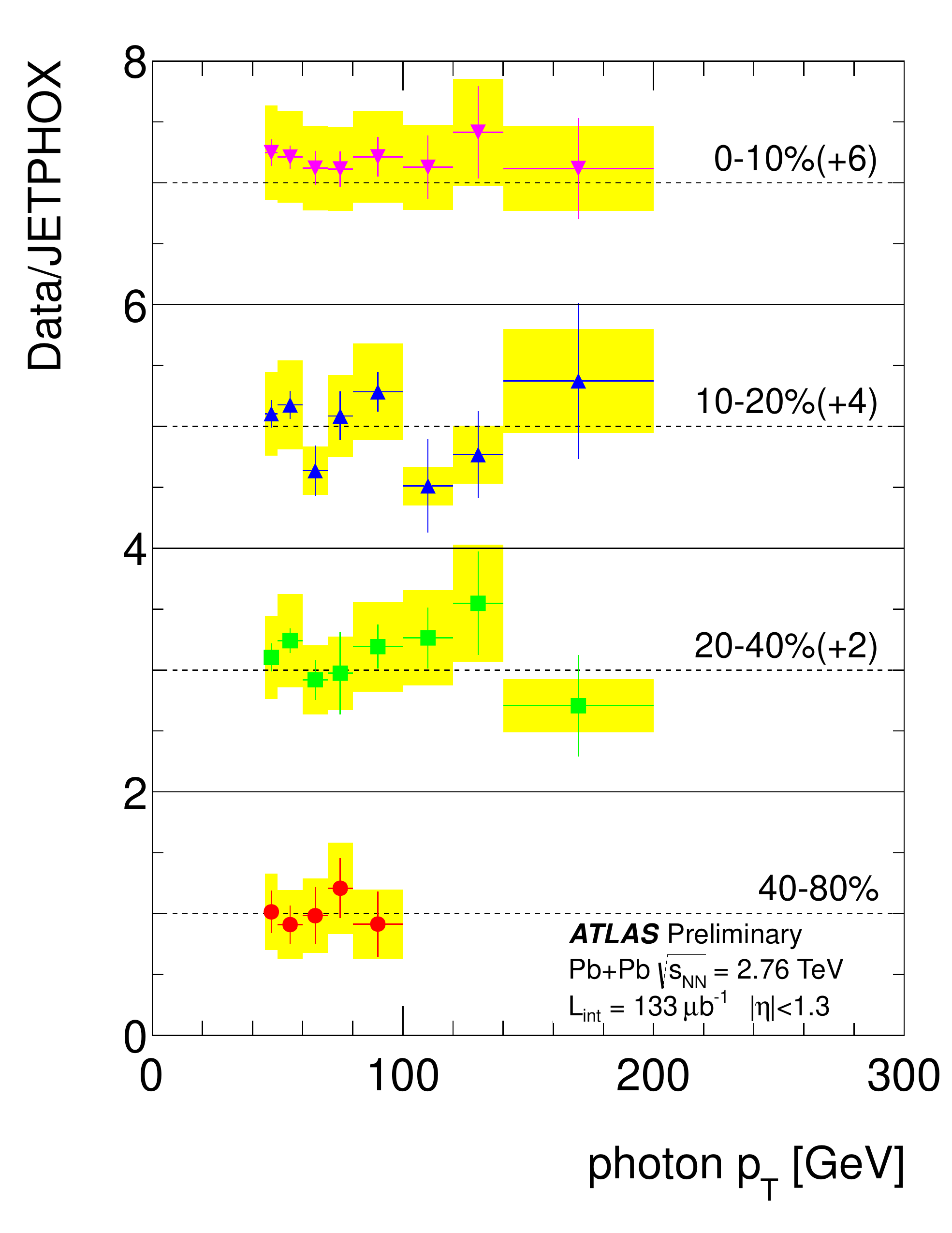}
\caption{
(left) Efficiency corrected yields
of prompt photons in $|\eta|<1.3$ using HI tight cuts, isolation cone radius \Riso\ = 0.3 and isolation energy of 6 GeV~\cite{HIPhoton}.  
Statistical errors are shown by the error bars.
Systematic uncertainties on the photon yields are combined and shown by the yellow bands.
(right) Efficiency corrected yields of prompt photons in $|\eta|<1.3$ 
using HI tight cuts, isolation cone radius \Riso\ = 0.3 and isolation energy of 6 GeV, 
divided by JETPHOX 1.3 predictions, which implement the same isolation selection.
Statistical errors are shown by the error bars.
Systematic uncertainties on the photon yields are combined and shown by the yellow bands.
\label{figure:yields}
}
\end{center}
\end{figure}

The photon yield is defined as
$\frac{1}{\Nevt}\frac{dN_{\gamma}}{d\pT} \left( \pT, c \right) = \frac{\Nsig_A }{\efftot\ \times N_{\mathrm{evt}} \times \Delta \pT}$
where $\Nsig_A$ is the background-subtracted yield in region A, $\efftot$ is the abovementioned total
efficiency, $N_{\mathrm{evt}}$ is the number of events in 
the centrality bin $c$, $\Delta \pT$ is the width of the
transverse momentum interval.
The final yields 
are shown 
divided by the mean nuclear thickness  $\meanTAA$ (which scales as the number of binary collisions)
as a function of photon $\pT$ in Figure~\ref{figure:yields}.
While the most peripheral interval (40-80\%) is unscaled, each subsequent 
centrality interval is scaled up by a factor of 10 for visibility.
Also included are CMS data~\cite{Chatrchyan:2012vq} for 0-10\%
central heavy ion collisions,
$|\eta|<1.44$ (a 10\% larger interval in $\eta$), 
and an isolation condition of at most 5 GeV transverse energy in a cone of radius $\Riso = 0.4$,
superimposed on the most central bin, as well as CMS $pp$ data~\cite{Chatrchyan:2012vq}
from $\sqrt{s}=2.76$ TeV, superimposed on the most peripheral bin.
The systematic uncertainties on the yield are indicated by yellow bands, and the uncertainties on the mean nuclear thickness are
tabulated in Ref.~\cite{HIPhoton}.
The ratio of the data to JETPHOX 1.3.0~\cite{Catani:2002ny,Aurenche:2006vj} is shown in the right panel of Figure~\ref{figure:yields}. 
It is observed that, within the statistical and systematic uncertainties, 
the data are consistent with the predictions in all centrality intervals.

\section{Conclusion}
Yields of isolated prompt photons in lead-lead collisions have been measured as a function of collision centrality
in a kinematic range of photon $\pT = 45-200$ GeV and for $|\eta|<1.3$.
Photons have been reconstructed using the large-acceptance, longitudinally segmented ATLAS electromagnetic calorimeter,
after subtraction of the average background, event-by-event.  
After scaling the yields by the mean nuclear thickness $\meanTAA$, 
they are observed to be constant as a function of centrality within experimental uncertainties, 
implying a linear scaling with the number of binary collisions.  
The scaled yields, as a function of $\pT$, are found to be in good agreement with 
next-to-leading order pQCD calculations, as implemented in JETPHOX 1.3.0.

\bibliographystyle{elsarticle-num}
\bibliography{<your-bib-database>}

\begin{thebibliography}{99}
\bibitem{Aad:2010bu}  ATLAS Collaboration,  Phys.\ Rev.\ Lett.\  {\bf 105}, 252303 (2010).
\bibitem{Wang:1996yh} X.~-N.~Wang, Z.~Huang and I.~Sarcevic,  Phys.\ Rev.\ Lett.\  {\bf 77}, 231 (1996). 
\bibitem{Adler:2005ig}   S.~S.~Adler {\it et al.}  [PHENIX Collaboration],  Phys.\ Rev.\ Lett.\  {\bf 94}, 232301 (2005). 
\bibitem{Chatrchyan:2012vq}  CMS Collaboration,  Phys.\ Lett.\ B {\bf 710}, 256 (2012). 
\bibitem{HIPhoton} ATLAS Collaboration, ATLAS-CONF-2012-051 (2012).
\bibitem{Aad:2008zzm} ATLAS Collaboration,  JINST {\bf 3}, S08003 (2008).
\bibitem{:2012is} 
  ATLAS Collaboration,
  arXiv:1208.1967 [hep-ex].
\bibitem{Aad:2009wy}  ATLAS Collaboration,  arXiv:0901.0512 [hep-ex].
\bibitem{ATLASlumi} \url{https://twiki.cern.ch/twiki/bin/view/AtlasPublic/LuminosityPublicResults#2011_Pb_Pb_Heavy_Ion_Collisions}
\bibitem{Sjostrand:2006za} T.~Sjostrand, S.~Mrenna and P.~Z.~Skands, JHEP {\bf 0605}, 026 (2006).
\bibitem{Wang:1991hta} X.-N. Wang and M. Gyulassy, Phys. Rev. {\bf D44}, 3501 (1991).
\bibitem{Agostinelli:2002hh} S.~Agostinelli {\it et al.}  [GEANT4 Collaboration], Nucl.\ Instrum.\ Meth.\  A {\bf 506}, 250 (2003).
\bibitem{ATLAS:2011ah} ATLAS Collaboration,  Phys.\ Lett.\ B {\bf 707}, 330 (2012). 
\bibitem{Aad:2010sp} ATLAS Collaboration,   Phys.\ Rev.\ D {\bf 83}, 052005 (2011).
\bibitem{Aad:2011tw} ATLAS Collaboration,  Phys.\ Lett.\ B {\bf 706}, 150 (2011).
\bibitem{Catani:2002ny}  S.~Catani, M.~Fontannaz, J.~P.~Guillet and E.~Pilon,  JHEP {\bf 0205}, 028 (2002). 
\bibitem{Aurenche:2006vj}  P.~Aurenche, M.~Fontannaz, J.~-P.~Guillet, E.~Pilon and M.~Werlen,  Phys.\ Rev.\ D {\bf 73}, 094007 (2006). 
\end{thebibliography}

\end{document}